\documentclass[aps,prl,showpacs,twocolumn,groupedaddress,floatfix]{revtex4}

\usepackage{graphics}

\begin{document}


\title{A New Limit on Signals of Lorentz Violation in Electrodynamics}


\author{J. A. Lipa}
\email[]{jlipa@stanford.edu}
\author{J. A. Nissen}
\author{S. Wang}
\author{D. A. Stricker}
\author{D. Avaloff}

\affiliation{Physics Department, Stanford University, Stanford, CA 94305, USA}

\date{\today}

\begin{abstract}
We describe the results of an experiment to test for spacetime
anisotropy terms that might exist from Lorentz violations. The
apparatus consists of a  pair of cylindrical superconducting cavity-stabilized
oscillators operating in the TM$_{010}$ mode with one axis east-west
and the other vertical. Spatial anisotropy is detected by monitoring
the beat frequency at the sidereal rate and its first harmonic.
We see no anisotropy to a part in
$10^{13}$. This puts a comparable bound on four linear combinations of
parameters in the general Standard Model extension, and a weaker bound 
of $<4 \times 10^{-9}$ on three others.
\end{abstract}


\maketitle



\def\om{\omega}
\def\fr#1#2{{{#1} \over {#2}}}

Tests of spacetime anisotropy are generally divided into two main
classes, one involving angle dependent effects and the other absolute
velocity effects. Following Robertson \cite{Robertson}, a treatment of 
potential Lorentz invariance violations involving idealized rods and
clocks was developed by Mansouri and Sexl \cite{Mansouri} who
considered the possibility of an anisotropic propagation velocity of
light relative to a preferred frame. In this model, if a laboratory is 
assumed to be moving with a velocity $v$ at an angle $\theta$
relative to the axis of a preferred frame, the speed of light as a
function of $\theta$ and $v$ is given by
\begin{eqnarray}
\frac{c{(\theta,v)}}{c} = 1 + \left({\frac{1}{2}} - \beta +
\delta\right)\left(\frac{v}{c}\right)^2\sin^2\theta \nonumber\\
+ (\beta - \alpha -1) \left(\frac{v}{c}\right)^2 \label{speedc}
\end{eqnarray}
where $\alpha$ is the time dilation parameter, $\beta$ is the Lorentz 
contraction parameter,and $\delta$ tests for transverse contraction.
In special relativity the last two terms on the right hand side of the 
equation are zero.  Classical Michelson-Morley experiments attempt to
measure the amplitude of the $\theta$-dependent term, while
Kennedy-Thorndyke experiments set limits on the amplitude of the 
$\theta$-independent term.  While useful to help categorize
experiments, this approach has a number of limitations. For example,
it fails to include effects on the measurement system itself and it
does not take into account the full range of anisotropies allowable in nature.
\def\om{\omega}
\def\fr#1#2{{{#1} \over {#2}}}

Recently Kosteleck\'y and Mewes \cite{Kostelecky} (KM) have pointed
out that in the Standard Model Extension (SME) that describes general
Lorentz violations \cite{Colladay}, additional terms may exist which 
show signatures different from those expressed in Eq. (\ref{speedc}). 
In particular, $\sin\theta$ and $\sin2\theta$ terms may exist
independent of $v$ which could be detectable in the experiments, as
well as terms first order in $v/c$. No systematic search for these 
terms appears to have been undertaken 
in the optical experiments conducted until now,
although a number of tests have been performed with fermions and
with astrophysical sources \cite{ref3}. Also, experiments involving the
Earth's rotation 
typically use co-rotating frequency references which complicates the 
analysis.  A simple configuration that can be analyzed easily consists 
of a pair of cylindrical microwave cavity resonators operating on
radial modes with their axes aligned in the east-west direction and 
optimally at $45^{\circ}$ to the Earth's axis, as indicated in 
Fig. 1. This apparatus will in general have a different sensitivity 
to the coefficients of the Lorentz violating terms than an optical 
cavity experiment because of the radial nature of the wave motions involved.
Each cavity will provide its own fractional offset signal $\delta \nu/ \nu$
from its unperturbed frequency.
The beat signal from such a pair,
$\Delta \nu/ \nu = \delta \nu_1/\nu - \delta \nu_2/\nu$, takes the general form
\begin{eqnarray}
\frac{\Delta \nu}{\nu} &=&
{\cal A}_{s}\sin\om t
+{\cal A}_{c}\cos\om t
\nonumber \\ &&
+{\cal B}_{s}\sin2\om t
+{\cal B}_{c}\cos2\om t + {\cal C} \label{beat}
\end{eqnarray}
where the coefficients are linear combinations of potential Lorentz 
violating terms in the SME and $2\pi/\omega$ is the Earth's sidereal
period.  The term ${\cal C}$ has an annual variation that can be 
neglected here.
In zero and first order of $v/c$ the quantities ${\cal A}$ and ${\cal B}$ contain 
exclusively SME terms, while higher order terms would of course
include the more traditional effects described by Eq. (\ref{speedc}). 
The cavity which is oriented in the east-west direction is maximally sensitive
to the second harmonic terms in Eq. (2), while the cavity oriented $45^{\circ}$ 
to the Earth's axis is maximally sensitive to the first harmonic terms 
\cite{Kostelecky_52}. 
A search for the lower order terms would probe for new physics that 
might for example correspond to residual effects left over from the 
birth of the universe. Because of the extreme sensitivity of modern cavity 
resonators and clocks it is possible to put useful bounds on such 
possibilities. 

When limited to the photon sector the Lagrangian describing
the SME can be written in the form \cite{noteref3}
\begin{eqnarray}
L_{photon} =  -{\frac{1}{4}} F_{\mu\nu}F^{\mu\nu} - {\frac{1}{4}}
(k_F)_{\kappa\lambda\mu\nu} F^{\kappa\lambda}F^{\mu\nu}\nonumber\\ 
+ {\frac{1}{2}}(k_{AF})^\kappa \epsilon_{\kappa\lambda\mu\nu}A^\lambda F^{\mu\nu} \label{lagrange}
\end{eqnarray}
Here $A_\mu$ are the photon fields, 
$F_{\mu\nu} \equiv \partial_{\mu}A_{\nu} - \partial_{\nu}A_{\mu}$ 
and the coefficients $k_F$ and $k_{AF}$ control the magnitude of 
the Lorentz violations. Stringent limits exist on the size of the 
$k_{AF}$ term, but the CPT-even $k_F$ term is only partially
constrained \cite{Kostelecky}. KM define matrices 
$\tilde{\kappa}_{e+}$, $\tilde{\kappa}_{e-}$, $\tilde{\kappa}_{o+}$ 
and $\tilde{\kappa}_{o-}$ and  $\tilde{\kappa}_{tr}$ with elements that are
parity even and
parity odd combinations of the coefficients $k_{F}$.  
These matrices arise naturally in the
analogous situation of wave propagation in a homogeneous anisotropic 
medium. Astrophysical tests constrain $\tilde{\kappa}_{e+}$ and 
$\tilde{\kappa}_{o-}$ at the $10^{-32}$ level \cite{astro}, while the 
other matrices are currently only weakly constrained. These include
nine additional coefficients of $k_F$ of which eight are in principle accessible 
via the present experimental configuration. Of these, four contribute 
directly to a possible frequency shift and three at first order in 
$v/c$, leading to high sensitivity tests.
Detailed expressions relating the coefficients of Lorentz violation to those 
in Eq. (\ref{beat}) have been given by KM up to first order in $v/c$. 
\begin{figure}
\includegraphics{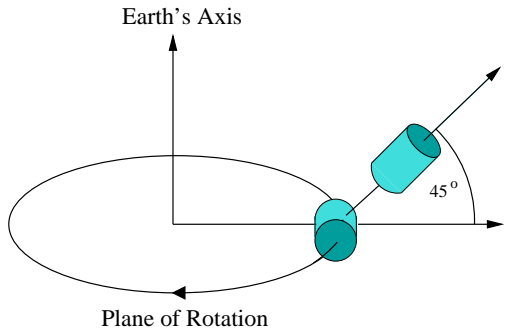}
\caption{\label{fig:epsart1}Ideal arrangement of two microwave
cavities relative to the Earth's rotational axis which maximizes 
sensitivity to sidereal and twice-sidereal variations in the beat frequency.
Actual tilt angle is the latitude of the laboratory.}
\end{figure}

In our experiment we compare the frequencies of
two cylindrical superconducting microwave cavities
operating in the TM$_{010}$ mode \cite{Stein}
which gives sensitivity to the velocity of light in radial directions. One 
cavity has its axis oriented to the local vertical
at our latitude of $37.4^{\circ}$ while the other axis is oriented to the local
horizontal in the east-west direction. The cavities are made
of niobium and are operated at 
about 1.4K in conventional helium Dewars. Microwave synthesizers are 
locked to the 8.6 GHz modes of the cavities using Pound frequency 
discrimination systems, and the difference frequency is mixed with an 
intermediate frequency oscillator to produce a beat signal in the 
20 - 30 Hz range. Data was collected at irregular intervals over a 
98-day period during a development program for a related experiment 
to be performed in space \cite{Sumo}. Nine records were obtained, each 
corresponding to a continuous segment of data at least 24 hrs long. 
Frequency sampling was at one second intervals, with averaging of 
100 second segments before curve fitting was performed.  An example of 
one of the records is shown in Fig. 2(a).  Typically the records were 
collected after some other form of testing on the apparatus was
completed. This situation leads to arbitrary offsets of up to a few Hz 
between the records. Also, mechanical disturbances occasionally gave
rise to a perceptible drift of the beat frequency amounting to a few
mHz per day. We therefore fitted each record with the function
%
\begin{eqnarray}
\frac{\Delta \nu}{\nu}  =  \nu_0 + \nu_1t + {\cal A}_{s}\sin(\omega t) + 
{\cal A}_{c}\cos(\omega t)\nonumber\\
+ {\cal B}_{s}\sin(2\omega t) + {\cal B}_{c}\cos(2\omega t)\label{fitbeat}
\end{eqnarray}
%
where $\nu_0$ and $\nu_1$ were additional free parameters. The
residuals from the fit to the record in Fig. 2(a) are shown in Fig. 2(b).  
The values obtained for the coefficients of the sinusoidal terms are
listed in Table I along with the day of the record \cite{notetime}.
It can be seen that the amplitudes ${\cal A}_{s}$, ${\cal A}_{c}$, ${\cal B}_{s}$ 
and ${\cal B}_{c}$ are
in the low $10^{-13}-10^{-14}$ range, with no obvious trend with time.
XY plots of the sine and cosine coefficients are shown in Figs. 3(a) and 3(b).  
To test for an alignment of the observed signals with inertial space
we also made the XY plots after phase shifting the sine and cosine 
coefficients to sidereal time, obtaining the results in Figs. 3(c) and 3(d). 
For these plots the time origin was also shifted to the 2002 vernal equinox.
No significant reduction of the scatter is evident.  
%
\begin{figure}
\includegraphics{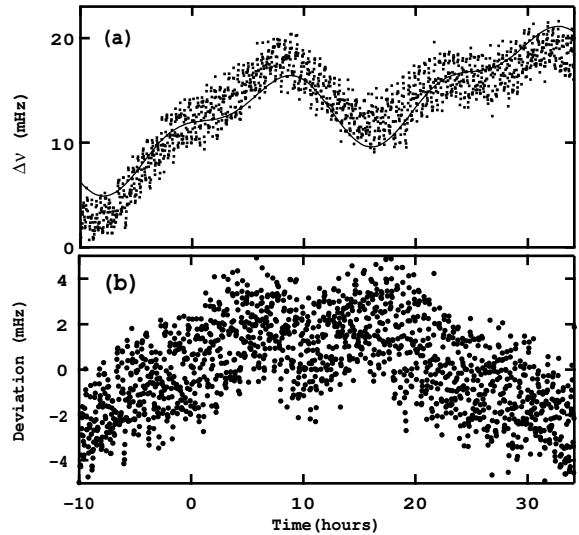}
\caption{\label{fig:epsart2}(a) Example of a beat frequency record $\Delta \nu $ as a function of time.
$\Delta \nu $ is measured from 26 Hz. Line shows best fit with Eq. (\ref{fitbeat}). 
(b) Residuals of data in (a) after subtraction of best fit.} 
\end{figure}
%

A study of the behavior of the apparatus at other times disclosed a high 
sensitivity of the signal from one of the
cavities to tilt which appears to be the dominant limit to
the experiment.  A second effect was the stability of the temperature of the cavity.
This was 
controlled to within $\pm 5\times 10^{-6}$ K using a germanium
resistance thermometer mounted on the
cavity and a servoed heater. The
dominant source of temperature fluctuation was the 1.4 K cooling
system. A proportional-integral temperature controller was used, but 
thermal gradients within the cavity assembly could cause some
undetected coupling with room temperature. A ratio transformer bridge 
was used for the germanium thermometer with a reference resistor at 
1.4 K. This configuration typically gives stabilities of better 
than 1 $\mu$K.  The temperature coefficient of the cavity frequency was 
-28 Hz/K which would imply frequency offsets on the order of $\pm 0.15$ mHz, 
but this could be amplified by thermal gradient effects. With servo 
powers of the order $10^{-5}$ W, temperature differences of as much as 
$5\times10^{-5}$ K would be expected in the cavity support structure.  
A number of correlation studies were performed but only a modest
reduction of the amplitudes in Table I was obtained.  Discussion of
this aspect of the analysis is lengthy and will be presented elsewhere.
We suspect that the signal amplitudes in Table I are dominated by mechanical
effects in the low temperature apparatus.  
%
\begin{table}
\caption{\label{tab:table1}Coefficients of sinusoidal terms from best fits 
to the raw data \cite{notetime}
with Eq. (\ref{fitbeat}). Uncertainties in the coefficients from the fit
are given in parentheses.}
\begin{ruledtabular}
\begin{tabular}{ccccc}
 Day&${\cal A}_{s}\times 10^{13}$&${\cal A}_{c}\times 10^{13}$&${\cal B}_{s}\times 10^{13}$&${\cal B}_{c}\times 10^{13}$\\
\hline
1 &  0.731 (0.04) &  2.520 (0.04) & -0.216 (0.04) & -0.204 (0.04) \\
3 & -0.081 (0.03) &  1.189 (0.03) & -1.680 (0.03) &  0.605 (0.03) \\
18 &  3.699 (0.12) &  0.368 (0.12) & -1.817 (0.11) &  0.691 (0.12) \\
26 &  2.286 (0.07) &  0.108 (0.07) & -0.950 (0.07) &  0.459 (0.07) \\
59 &  2.503 (0.08) & -0.697 (0.09) & -1.347 (0.08) &  1.101 (0.08) \\
78 & -0.329 (0.10) &  1.776 (0.13) &  0.535 (0.10) & -0.457 (0.11) \\
80 &  1.006 (0.06) &  0.515 (0.06) & -0.156 (0.06) & -0.135 (0.06) \\
95 & -0.809 (0.06) &  0.107 (0.06) & -0.212 (0.06) &  0.145 (0.06) \\
98 & -0.306 (0.04) & -1.336 (0.04) &  0.823 (0.04) & -0.558 (0.04) \\
\end{tabular}
\end{ruledtabular}
\end{table}

From the plots in Fig. 3 it seems reasonable to conclude that there is
no significant evidence for an inertially oriented frequency shift in 
our experiment. Treating the variation of the observed signal
amplitudes as locally generated "noise" we can average the data in
each direction and derive bounds on any signal.  Using the
coefficients from Figs. 3(c) and 3(d) we obtain  
$\bar{\cal A}_{c} = -8.5\pm 10.4 \times 10^{-14}$, 
$\bar{\cal A}_{s} =  4.2\pm 8.8 \times 10^{-14}$, 
$\bar{\cal B}_{c} = -2.0\pm 4.3 \times 10^{-14}$ and  
$\bar{\cal B}_{s} =  5.8\pm 5.9 \times 10^{-14}$  where the 
errors correspond to the statistical 2$\sigma$ level. Because of
the likely presence of 
unmodeled systematic effects and the small number of records, we consider the confidence
level in these results to be closer to the 60\% or 1$\sigma$ level.  
The results imply that certain linear combinations of the $k_F$
coefficients are constrained at the $10^{-13}$ level. 
For example, neglecting contributions of order $v/c$ and higher, the ${\cal A}_{s}$ term can be
written as 
${\cal A}_{s}^0  = 1/4 \sin2 \chi (3\tilde{\kappa}_{e+} + \tilde{\kappa}_{e-})^{YZ}$ where $\chi$ is the
colatitude of the experiment \cite{notexyz}. Setting $(\tilde{\kappa}_{e+})^{YZ} = 0$ on
the basis of the extremely 
tight astrophysical bound $< 10^{-32}$, 
it is easy to show that ${\cal A}_{s}^0 = (1/16) \sin2\chi [(k_F)^{ZXYX} + (k_F)^{XZXY} - (k_F)^{ZXXY}
 - (k_F)^{XZYX} - 4(k_F)^{0Y0Z}]$, which reduces to $\sin2\chi[(k_F)^{XYXZ} - (k_F)^{0Y0Z}]/4$.
Similar relations can be derived for the other amplitudes in Eq. (\ref{beat}). 
Alternatively, the experiment can be viewed
as setting the following bounds on elements of the $(\tilde{\kappa}_{e-})$ matrix:
$(\tilde{\kappa}_{e-})^{YZ} < 1.7 \pm 3.6 \times 10^{-13}$;
$(\tilde{\kappa}_{e-})^{XZ} < -3.5 \pm 4.3 \times 10^{-13}$;
$(\tilde{\kappa}_{e-})^{XY} < 1.4 \pm 1.4 \times 10^{-13}$;
$[(\tilde{\kappa}_{e-})^{XX} - (\tilde{\kappa}_{e-})^{YY}] < -1.0 \pm 2.1 \times 10^{-13}$.

\begin{figure}
\includegraphics{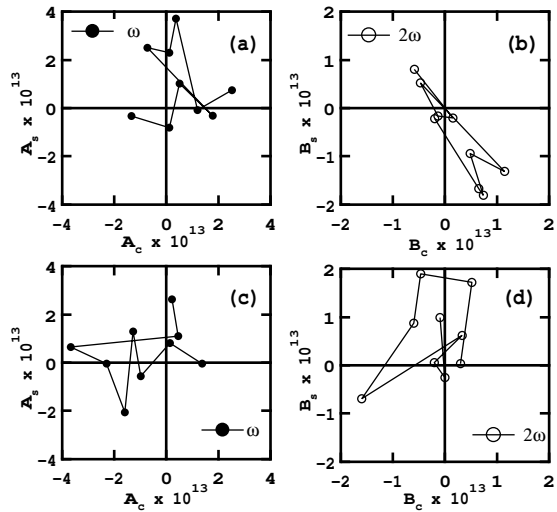}
\caption{\label{fig:epsart3}XY plots of best fit coefficients with Eq. (\ref{fitbeat}) for all records. 
 (a) ${\cal A}_{s}$ vs. ${\cal A}_{c}$, and (b) ${\cal B}_{s}$ vs. ${\cal B}_{c}$ using solar time;
 (c) and (d): same as (a) and (b) but with sidereal phase shifts added to align results to inertial
 space. Lines link the data points in time sequence.}
\end{figure}
%

As described by KM,
three more coefficients 
are introduced at the level $v/c$, where $v$ is the velocity of the Earth around the 
Sun. The additional term in ${\cal A}_{s}$ can be written as
\begin{eqnarray}
{\cal A}_{s}^1 =  \frac{v}{4c} \sin2\chi \cos\Omega T[ \sin\eta(\tilde{\kappa}_{o+})^{ZX} 
- \cos\eta(\tilde{\kappa}_{o+})^{YX}] \label{AS}
\end{eqnarray}
where $2\pi/\Omega$ is the orbital period of the Earth, $T$ is the time measured from the 
vernal equinox and $\eta$ is the angle between the Earth's orbital and 
equatorial planes. Evaluating this expression for our situation we obtain  
${\cal A}_{s}^1 = 9.75\times 10^{-6} [(\tilde{\kappa}_{o+})^{YX} - 0.432(\tilde{\kappa}_{o+})^{ZX}]$
where we have used the midvalue of $T$ over our data collection period.
Clearly this expression is dependent on the duration of the experiment.
In conjunction with ${\cal A}_{s}^0$ this term is also bounded at the level
of $4.2\times 10^{-14}$, implying a constraint 
on the term inside the square brackets of $< 4.0 \pm 8.4 \times 10^{-9}$, assuming no cancellation 
between the terms \cite{notesigma}. Similar expressions can be derived for the other 
coefficients in Eq. (\ref{fitbeat}). For clarity, the entire
set of constraints obtained from
the measurements is given in Table II.
A more direct bound on the components of ${\cal A}_{s}^1$ could be 
obtained by extending the data gathering period to a larger fraction of a year 
when it would become reasonable to include the annual modulation terms in the fit 
to the data. 
At the present level, the experiment sets bounds of $ < 4 \times 10^{-9}$
on four expressions of the type in square brackets in Eq. (\ref{AS}).
We note that by mixing the experimental bounds, a cleaner separation of the
$\tilde{\kappa}_{o+}$ components can be obtained.
%
\begin{table}
\caption{\label{tab:table2}Experimental constraints on coefficients in the SME assuming no
cancellation effects.}
\begin{ruledtabular}
\begin{tabular}{cc}
 Constrained Quantity & Bound\\
\hline
$(\tilde{\kappa}_{e-})^{YZ}$ &  $ 1.7 \pm 3.6 \times 10^{-13}$ \\
$(\tilde{\kappa}_{e-})^{XZ}$ &  $ -3.5 \pm 4.3 \times 10^{-13}$ \\
$(\tilde{\kappa}_{e-})^{XY}$ &  $ 1.4 \pm 1.4 \times 10^{-13}$ \\
$(\tilde{\kappa}_{e-})^{XX} - (\tilde{\kappa}_{e-})^{YY}$ &  $ -1.0 \pm 2.1\times 10^{-13}$\\
$(\tilde{\kappa}_{o+})^{YX} - 0.432(\tilde{\kappa}_{o+})^{ZX}$ & $ 4.0 \pm 8.4 \times 10^{-9}$\\
$(\tilde{\kappa}_{o+})^{XY} - 0.209(\tilde{\kappa}_{o+})^{YZ}$ & $ 4.0 \pm 4.9 \times 10^{-9}$\\
$(\tilde{\kappa}_{o+})^{XZ} - 0.484(\tilde{\kappa}_{o+})^{YZ}$ & $ 1.6 \pm 1.7 \times 10^{-9}$\\
$(\tilde{\kappa}_{o+})^{YZ} + 0.484(\tilde{\kappa}_{o+})^{XZ}$ & $ 0.6 \pm 1.9 \times 10^{-9}$\\
\end{tabular}
\end{ruledtabular}
\end{table}

By restricting the Lagrangian in Eq. (\ref{lagrange}) to the photon sector, the model 
omits a range of potential
effects from the material that makes up the apparatus.  Within the full SME these
possibilities lead to considerable complexity, but KM argue that complete cancellation of
the photon sector effects is improbable due to the complexity of the forces involved.
In very recent work M\"uller \textit{et al}. \cite{Muller} have considered
these effects and find a small enhancement of the photon effects in the case of ionic crystalline
materials.

In summary, we have set bounds at the $10^{-13}$ level on four combinations of parameters in the
SME, and bounds at the $10^{-9}$ level on four others. These bounds now constrain seven of the
nine unknown coefficients $k_F$ in the model.
We note that in a space-based version of this experiment \cite{Sumo}, substantially
greater sensitivity could be
achieved, perhaps approaching the  $10^{-17}$ level for the four primary bounds.

We wish to thank J. Turneaure and S. Buchman for support in the early phases
of the project, the NASA Office of Biological and Physical Research
for its support with Grants No. NAG3-1940 
and No. NAG8-1439, and the Jet Propulsion Laboratory for Contract No. 1203716.


\begin{thebibliography}{99}
\bibitem{ref3}This situation has been discussed in some detail in ref. 3.
\bibitem{Kostelecky_52}The sensitivity of the experiment to cavity orientation 
is described to first order in $v/c$ by Eq. (52) of Ref. 3.
\bibitem{noteref3}We use the notation of ref. 3 where applicable.
\bibitem{notesigma}If a statistically significant signal was observed, 
some progress could be made untangling the effects of the various contributions.
\end{thebibliography}
\end{document}